\documentstyle[preprint,aps,psfig]{revtex}
\begin{document}
\draft \title{ Liquefaction of a horizontally vibrated granular bed:
  Friction, Dilation and Segregation}
\author{ Sarath G. K. Tennakoon and R. P. Behringer}
\address{Department of Physics and Center for Nonlinear and Complex 
Systems, Duke University, Durham, North Carolina 27708-0305 }
\date{\today}
\maketitle

\begin{abstract}
  We present experimental observations of the onset of flow
  (liquefaction) for horizontally vibrated granular materials.  As the
  acceleration $\Gamma$ increases above $\Gamma^*$, the top layer of
  granular material liquefies, while the remainder of the layer moves
  with the shaker in solid body motion.  With increasing $\Gamma$,
  more of the layer becomes fluidized.  The initial bifurcation is
  backward, and the amount of hysteresis depends mainly on frictional
  properties of the granular media.  A small amount of fluidization by
  gas flow lifts the hysteresis.  Modest differences in the frictional
  properties of otherwise identical particles leads to rapid
  segregation.

\end{abstract}


Although granular materials are common in nature and industrial
applications, the complete understanding of their dynamical behavior
is still an open problem.  Consequently, the dynamics of granular
materials has attracted considerable interest in recent years (for
comprehensive reviews see reference\cite{behringer93} and citations
therein).  Granular materials can exhibit both fluid-like and
solid-like properties depending on the circumstances: they resist
shearing up to a point, but flow freely under strong enough shear or
at low enough density. These materials also display a number of
different dynamical states including liquefaction, heap formation and
convection under vibration, the spontaneous formation of stable
arches, segregation, a variety of density waves, stick-slip motion
during avalanches, etc.\cite{behringer93}

Much recent attention has been focused on the dynamics of vertically
vibrated granular materials\cite{behringer93}.  Although there have
been some
studies\cite{evesque_92,rosenkranz_97,pouliquen_97,evesque_97,jaeger_97},
much less is known about the corresponding dynamics of granular
materials subject to horizontal vibration, and of the existing work
much is very recent.  The aim of the present study is to explore the
dynamics of a horizontally vibrated systems, with a particular focus
on the transition to flow---sometimes referred to as liquefaction.  A
better understanding of this second case is of interest scientifically
because it gives insight into segregation phenomena and into the
competition between dilation and friction.  It is also of interest
because both horizontal and vertical vibration are commonly used in
industries as an aid to mixing, segregating and transporting granular
materials. Finally, soil liquefaction during earthquakes is a common
and destructive phenomena associated with horizontal shaking.

\begin{figure}[h]
\parbox{7.0in}{
\psfig{file=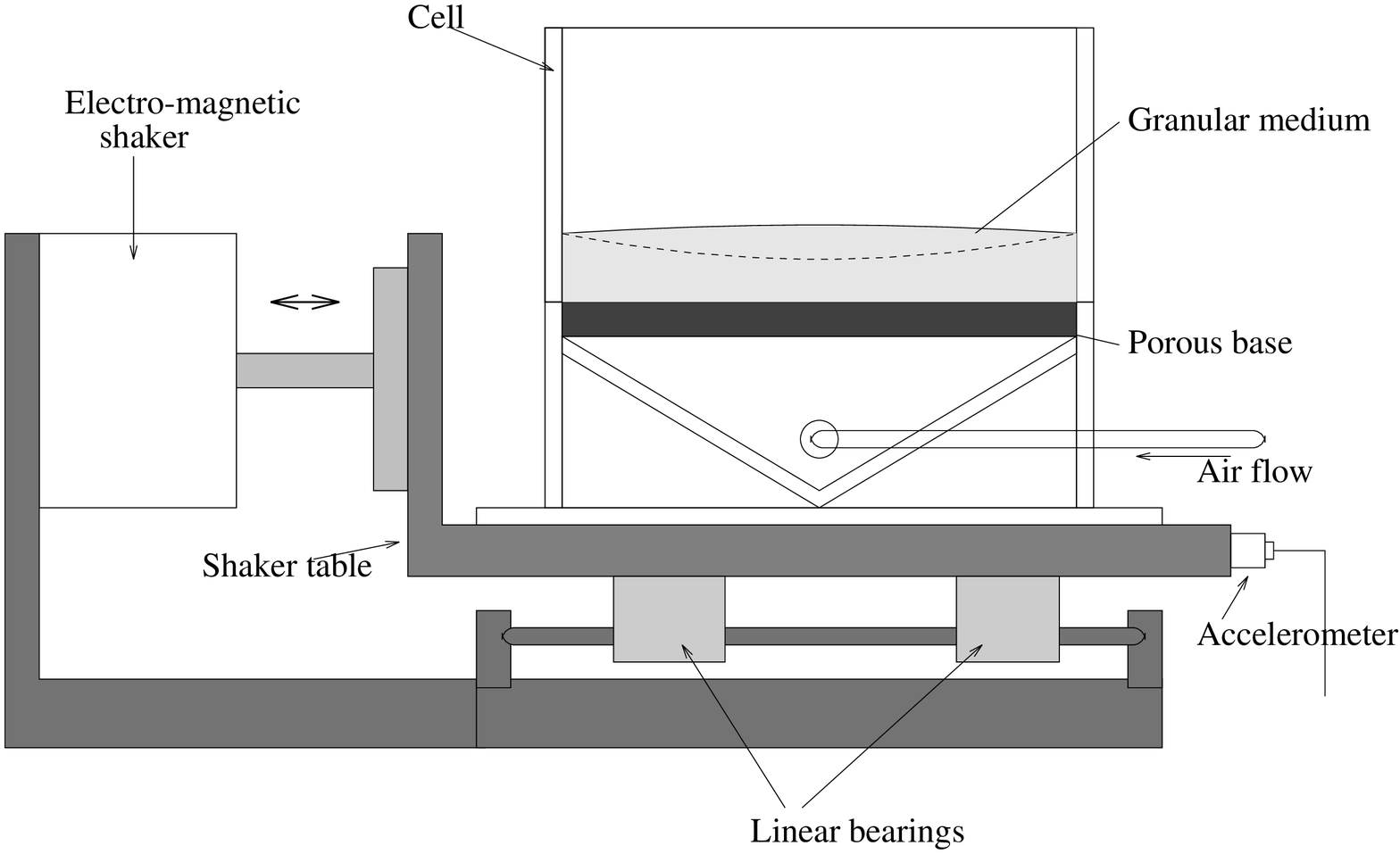,width=5.5in,rwidth=5.5in}
}
\caption{Schematic of the apparatus. The rectangular cell is made
  of Plexiglas, and is mounted on a another Plexiglas cell of the same
  cross-sectional dimensions which is attached to a small table.  The
  table is mounted on four linear bearings running on horizontal
  cylindrical guidance rods rigidly attached to a fixed bottom frame.
  The bottom section of the cell acts as a gas distributor.  An
  electro-mechanical vibrator, driven by a sinusoidal AC signal
  provides the horizontal driving.}
\label{fig:apparatus}
\end{figure}

Our experimental setup is shown in Fig ~\ref{fig:apparatus}.  The
heart of the experiment is a rectangular Plexiglas cell with
cross-sectional dimensions of $1.93\; cm$ by $12.1\; cm$. The base of
the cell is made of a porous medium (average pore size $50 \mu m $)
through which gas can flow in order to fluidize the granular medium.
This provides an independent control over the dilation of the
material.  The cell is mounted on a Plexiglas base of the same
cross-sectional dimensions as the cell, which is in turn mounted on a
movable table; the base acts as the gas distributor to the system.
The table is mounted on four linear bearings sliding on horizontal
shafts rigidly attached to the fixed bottom frame.  An
electro-mechanical actuator provides a sinusoidal drive of the form
$x\, = \, A \sin \omega t $ at frequencies, $\omega$, spanning of $3 -
15\; Hz$ and at amplitudes $A$, spanning $0 - 15mm$.  A calibrated PCB
accelerometer mounted on the moving table yields the acceleration of
the system.  This device indicates very little extraneous noise and
very nearly sinusoidal motion.

In a typical run, we observed the evolution of the system as $A$ was
increased from zero while keeping $\omega$ fixed.  We used several
types of approximately monodisperse granular materials, including
spherical glass beads, smooth Ottawa sand, and seived rough sand.  For
these experiments, there are several useful dimensionless dynamical
measures, including a dimensionless measure of the acceleration
$\Gamma\,=\, A {\omega}^{2}/g $ where $g$ is the acceleration due to
gravity.  We note that other parameters, such as $E = (A \omega)^2/g
d$, which are important in describing higher order phenomena in
vertically shaken materials (e.g. traveling waves\cite{pak_93},
coarsening\cite{vandoorn_97}) are not necessary to describe the onset
of flow in these experiments.  However, frictional properties are
imporant, and recent experiments\cite{pouliquen_97} also show that the
ratio $A/d$ may be important.

\begin{figure}[h]
\parbox{6.0in}{
\psfig{file=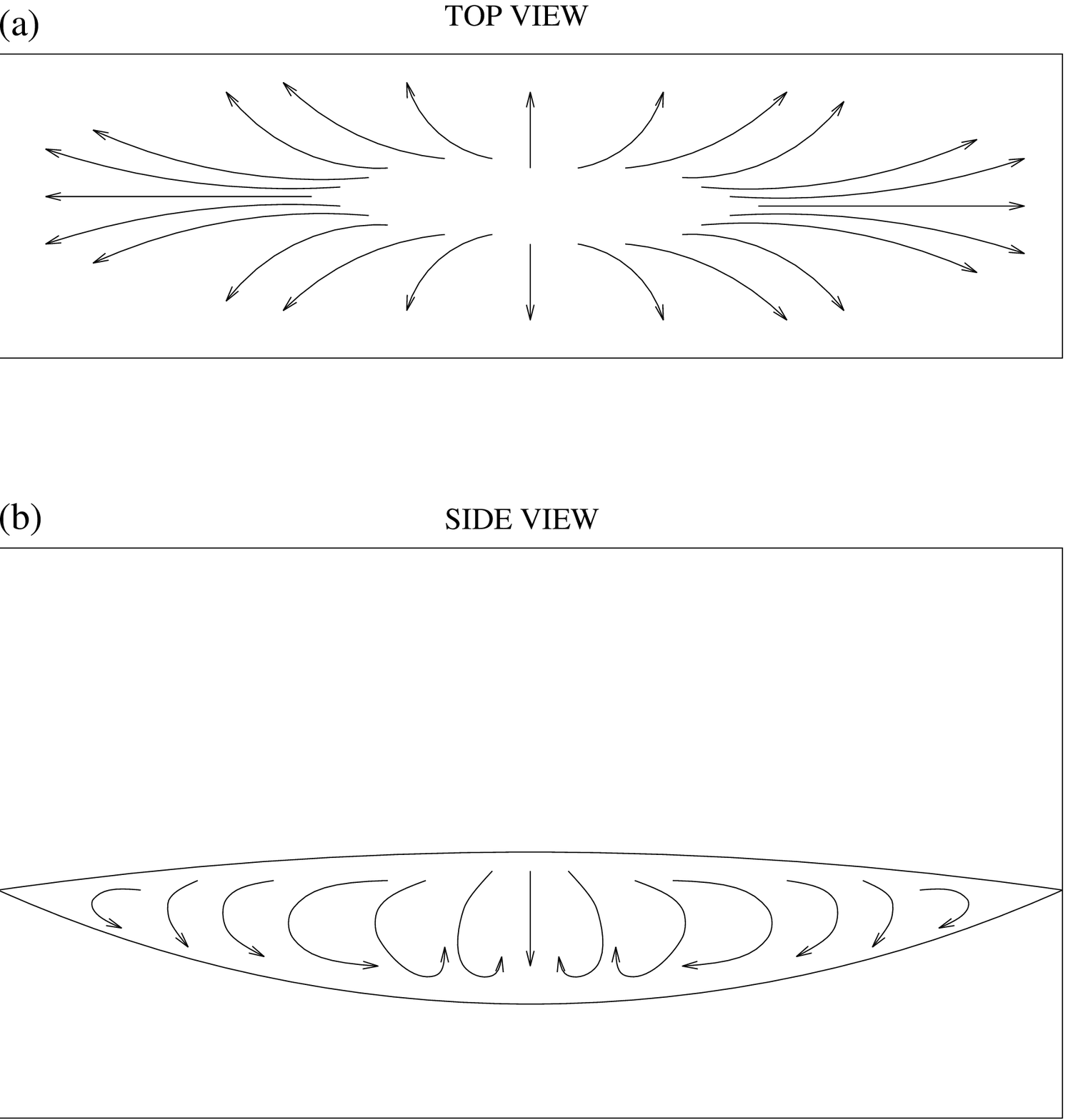,width=4.5in,rwidth=4.5in}
}
\caption{Sketch of convection flow lines in the liquified
  layer induced by horizontal shaking (a) as seen from the top, and
  (b) as seen from the side.  Grains rise in the middle of the cell,
  flow along the surface towards the side walls, and then sink at the
  wall boundaries.}
\label{fig:flowsketch}
\end{figure}

Fig ~\ref{fig:flowsketch} shows a sketch of convective flow lines
observed from the top and side for $\Gamma$ somewhat above onset. We
have obtained these images by coloring some of the particles and by
then following them over time.  (In this regard, considerable care
must be taken, since coloring the grains can change the surface
friction and lead to strong segregation, as discussed below.)  Both
lab-mounted and shaker-mounted cameras were used to observe the
system.  Grains rise up in the middle of the cell and flow along the
surface towards the side walls and then sink at the wall boundaries.
The top surface of the liquefied layer has a shape which is concave
down, and the bottom surface of this layer has a shape which is
concave upwards.  Thus, the thickness of the fluidized layer is
largest in the middle of the cell and smallest at the end walls, as
seen previously by Evesque\cite{evesque_92}.

A useful measure of the strength of the flow is then the thickness,
$H$, of the liquefied material in the middle of the cell.  We show
typical behavior for $H$ as a function of $\Gamma$ in Fig.
is hysteretic.  With increasing $\Gamma$, there is a well defined
transition to finite amplitude (i.e. finite $H$) flow at $\Gamma^*$.
If we then decrease $\Gamma$ below $\Gamma^*$ once flow has begun, the
thickness of the layer also decreases until $\Gamma$ reaches a
critical value $\Gamma_{c}$ where the relative motion completely
stops.  As $\Gamma$ is decreased from above towards $\Gamma_{c}$,
grains near the walls stop moving first, while grains in the middle
keep moving.  It is perhaps not surprising that the initial transition
to flow is hysteretic, since the onset of flow must occur by the
breaking of static friction, whereas once flow has begun, dynamical
friction is involved, assuming that grains remain in motion throughout
much of the shaking cycle.  In addition, the grains must dilate in
order to flow\cite{reynolds1885}, but once dilated, less energy is
likely to be required to sustain the flow.

Our observations show that the location of $\Gamma^{*}$ and
$\Gamma_{c}$ are reproducible for a given height of a particular
material. That is, different $\omega$'s or $A$'s yield the same
critical $\Gamma$'s for a given material and fill height.  Thus,
$\Gamma$ is the relevant control parameter, as opposed to say $E$ or
some other dynamical measure.

\begin{figure}[h]
\parbox{7.0in}{
\psfig{file=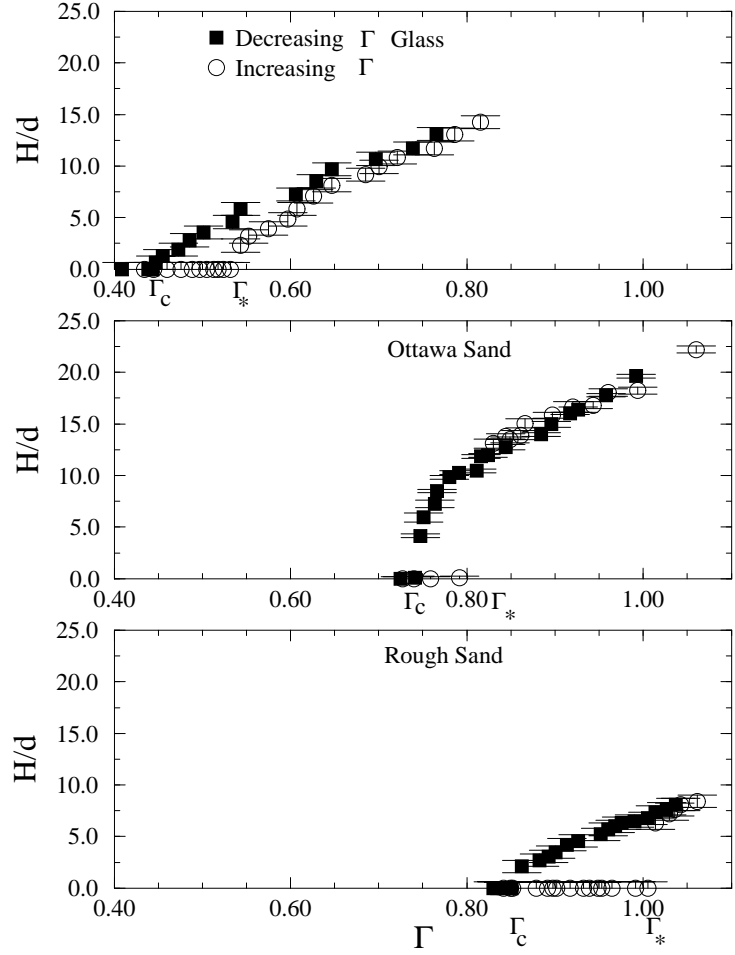,width=6.0in,rwidth=6.0in}
}
\caption{Thickness of the liquefied layer versus acceleration
  amplitude $\Gamma$ for (a) glass beads ($ d=0.6mm$), (b) smooth
  Ottawa sand ($d_{ave}=0.6mm$), and (c) rough sand ($d_{ave}=0.6mm$).
  In each of the figures, $\Gamma_{*}$ is the bifurcation point when
  $\Gamma$ is increased quasi-statically, and $\Gamma_{c}$ is the
  bifurcation point when $\Gamma$ is decreased quasi-statically.}
\label{fig:bifurcation}
\end{figure}

These $\Gamma$'s do depend on the physical properties of the material,
as shown in Fig ~\ref{fig:bifurcation}.  For instance, $\Gamma_c$ and
$\Gamma_*$ increase as the roughness of the granular materials
increases.  The same is true for the difference $(\Gamma^{*} -
\Gamma_{c})$, which is higher for rougher granular beds than for
smooth ones.  This can be attributed to two factors: first, the
ability of rough grains to roll is reduced because of the interlocking
of grains; and second, the effective macroscopic frictional forces
between grains and between grains and walls may also be higher for
rough grains.  These effects are in principle distinct, although in an
experiment, it may be difficult to distinguish between them.

\begin{figure}[h]
\parbox{6.0in}{
\psfig{file=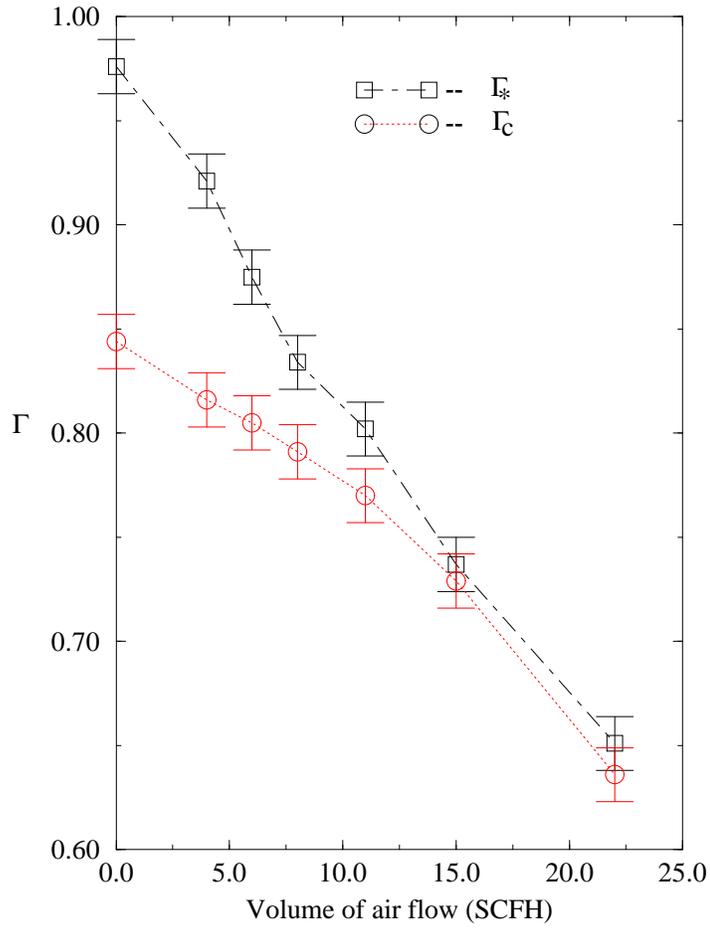,width=5.0in,rwidth=5.0in}
}
\caption{The bifurcation points $\Gamma^{*}$ and $\Gamma_{c}$ versus
  air flow rate for the rough sand.}
\label{fig:airflow}
\end{figure}

To obtain additional insight into the relative importance of dilatancy
and friction, we have fluidized the granular bed by passing air
through it, using the flow-controlled air supply (see
Fig.~\ref{fig:apparatus}), where the porous base of the cell acts as
the gas distributor.  The pressure gradient across the distributor is
more than $50\%$ of the total pressure gradient; hence, there is
reasonably even air distribution across the granular bed.
Fig~\ref{fig:airflow}, which presents $\Gamma^{*}$ and $\Gamma_{c}$ as
a function of air flow through the bed, indicates a very strong
dependence of these quantities on the air flow.  In particular, a
modest air flow reduces the critical $\Gamma$'s and ultimately
effectively removes the hysteresis in the initial transition.  A key
point is that the measured dilation of the bed due to the air flow is
very small; the maximum dilation for these experiments corresponds to
less than one granular layer for a bed of 45 layers, i.e. less than
$2\%$ dilation.  This small value of the dilation suggests that it is
the reduction in both static and dynamical friction which is most
important here rather than the unlocking effect which occurs for
larger dilations.  However, this conclusion must be made with some
caution, because it is possible to have motion in a granular material
along a localized shear band where the overall dilation of the whole
sample is also only a few percent.  In the experiments described here,
the dilation occurs over the whole sample (although not necessarily
completely uniformly).

Segregation by size and shape is a very common phenomenon observed in
granular systems and these experiments seem particularly sensitive to
this phenomenon.  We observed the usual segregation by size in flowing
layers of polydisperse grains.  The larger particles rose to the top
surface and accumulated near the surface and wall boundaries and the
small particles sank, mostly accumulating near the bottom of the
liquefied layer during the shaking.  However, even a modest difference
in the surface preparation of these materials was sufficient to cause
segregation--i.e. size differences are not necessary for segregation.
In particular, colored particles, which had a somewhat higher friction
coefficient than uncolored ones, tended to migrate to the upper
surface, where they accumulated next to the horizontal walls.  In the
case of size segregation for vertically shaken granular materials,
segregation is understood in terms of preferred downwards motion of
smaller particles throughout the medium\cite{rosato_87}, or in terms
of a dilated boundary layer which smaller but not larger particles can
penetrate\cite{knight_95}.  In the present case, the colored and
uncolored particles are virtually identical in size, so that an
explantation involving only friction-related mobility is necessary.
Mobility-related segregation occurs in conventional fluids (i.e.
separation of differenct species due to a temperature gradient), but
we are unaware of descriptions of segregation by mobility for granular
materials.  The segregation takes much more complicated formations at
high $\Gamma$ values where the granular convection dominates in the
liquefied layers.

To conclude, we have characterized the transition to flow, or
liquefaction, for horizontal shaking of granular materials.  In the
absence of additional fluidization by vertical gas flow, this
transition is hysteretic.  However, by applying a very modest vertical
gas flow which creates a dilation in the vertical of $\sim 2\%$, the
hysteresis is lifted, and the onset to flow occurs at lower $\Gamma$.
This suggests, at least tentatively, that it is the solid friction and
not the dilatancy which is responsible for the hysteresis.  In
addition, there is strong segregation of equal size particles if their
surface properties differ.  This phenomena is then qualitatively
different from the size-related segregation which has been reported
for vertical shaking.

Acknowledgments: This work was supported by the National Science
Foundation under Grant DMS95-04577, and by NASA under Grant NAG3
-1917.  RPB would like to thank the P.M.M.H-E.S.P.C.I for its
hospitality during the completion of this paper.

\end{document}